\documentclass[aps,prl,amsfonts,amssymb,twocolumn,groupedaddress]{revtex4}
\usepackage{graphicx}

\begin{document}

\title{Comment on
"Proximity effect in ultrafin Pb/Ag multilayers within the Cooper limit"
}
\author{M.V. Feigelman}
\affiliation{L.D.Landau Institute for Theoretical Physics, Moscow, Russia}

\maketitle

The paper~\cite{1} is concered with experimental data
on "proximity effect in ultrafin Pb/Ag multilayers within the Cooper limit".
While the authors do present some new experimental data,  their
presentation, to my point of view, is seriuosly deficient in  
i) adequate analysis of those data, and ii) in proper account of the
results obtained previously within the same field of research.

Concering item i), I have two different points to mention: 
a) While McMillan's formula for $T_c$ of strong-coupled
superconductor (Eq.(2) of~\cite{1})  is well-known, as well as the 
Cooper-limit formula for $T_c$ of S-N bilayer (Eq.(3) of ~\cite{1}), there
is no any reason to believe that {\it combination of both forumulae}
makes any physical sense. 
Indeed, what are the reasons to use in the
 McMillan's formula some average electron-phonon coupling 
$\lambda_{SN}$ while keeping the same constant Coulomb repulsion 
constant $\mu^*$ known for Pb ?  The Eq.(3) was derived within the BCS  
weak-coupling model of superconductivity, and its extension to
 strong-coupling case needs some additional work at least. 
The authors argue that
"electrons probe the entire sample" - why then the authors
assume they can neglect Coulomb replusion in the Ag layers ?
Without clear answers to these questions the "fit" used to describe
the obtained experimental results will stay extremely speculative.
b) The author's idea that bilayers studied can be described as being
within "Cooper limit" is based implicitely upon the assumption
that resistance of the S-N interface is sufficiently low
(it's trivial to see that in the case of low-conductive interface
the "Cooper limit" formula is not valid). Theory of
$T_c(\rho)$ dependence for thin S-N bilayer upon S-N interface 
resistance $\rho$ was presented in e.g. papers~\cite{bergman,hus,fom}.
While it is possible that samples studied in~\cite{1} do indeed belong
to the Cooper limit, this is to be checked carefully.

Now I proceed to the item ii). The authors claim they have shown
in this paper and in their previous paper\cite{2} that they
called "inverse proximity effect" - i.e. increase of $T_c$
of ultrafin S-N bilayers with increase of normal metal thickness $d_N$.
In fact, this effect was for the first time experimentally observed
more that 10 years before, in paper~\cite{LSD1}, and then studied 
in more details in~\cite{LSD2}. The authors of ~\cite{2} exactly 
reproduce the idea and experimental approach of~\cite{LSD1,LSD2} 
for the study of $T_c$ of ultrathin bilayers (just with another
choice for normal and superconductive metals), without any reference
to those papers. Detailed studies of $T_c$ behavior in multilayers
and bilayers can be found, e.g. even in earlier 
papers~\cite{beasley,bergman2}.

On theoretical side, the authors ignore the existence of
broadly accepted detailed microscopic theory of that the authors 
called "inverse proximity effect", cf.~\cite{Fin1,Fin2,Fin3} - not to 
mention previous papers on the same issue~\cite{fuku,kuroda}.  
In particular, in the review paper~\cite{Fin3} a detailed analysis
 of the experiments~\cite{LSD1,LSD2} is presented. {\it None of these
papers} is even mentioned in~\cite{1,2}. 

There exists however somethat new (at least, in comparison with
~\cite{LSD1,LSD2}) element in the experiments reported in
\cite{1,2}: apart from determination of $T_c$ via in-plane
 resistive measurements, the authors studied tunnelling conductance into the
bilayer, thus determining energy gap dependence upon thicknesses
of S and N layers. Theoretical results for the energy gap in 
S-N bilayers are published in e.g.~\cite{fom} within BCS mean-field model;
modification of tunnelling conductance in dirty S-N
films due to enchanced Coulomb interaction was discussed in~\cite{oreg97}.
 None of those papers is refered to in~\cite{1,2}.

\end{document}